\documentstyle[aps,prl,tighten,psfig]{revtex}
\begin{document}
\topmargin -.1cm

\title{
On surface energy of graphene and carbon nanoclusters}
\topmargin -.1cm

\author{Slava V. Rotkin}
\address{Beckman Institute, UIUC, 405 N.Mathews, Urbana, IL
61801, USA; \\
Ioffe Institute, 194021, St.Petersburg, Polytehnicheskaya
26, Russia.
\footnote{E-mail: rotkin@theory.ioffe.rssi.ru}}
\date{July 3, 2001}
\maketitle
\begin{abstract}
Theoretical study of graphite (graphene) edge is done. The most
stable edge orientation is calculated to be a zigzag [110] edge.
Possible applications of the result to the formation of
different graphitic structures are discussed.
\end{abstract}

%%%%%%\pacs{PACS: 81.05;61.48;31.10+R}

\section{Introduction}
\label{sec:1}

Recently, attention of physicists and chemists was recalled to
the problem of reconstruction of edges
of planar (and curvilinear)
graphite. Considerable amount of articles on
nanocarbons\cite{nanocarbons}, surmounted with recent results on
nanotubes\cite{shekhar} and graphite polyhedral
crystals\cite{gpc}, shows that this question is far from being
clear, despite intensive experimental and theoretical research
has been conducted.

The aim of this paper is to get theoretical understanding of
the formation of facets and
edges of graphite, graphene (monolayer of graphite) and small
carbon sheet fragments. I will use simple analytical approach
implementing the continuum energetics model\cite{mrs}
assumptions.
%
%%and discuss possible corrections to be done.
%
This approach bases on the count of dangling bonds along the
free edge of graphene. That gives an energy of the
facet\footnote{Henceforth I will use the term {\it facet} for
the 2D planar face of 3D graphite crystal and 1D straight edge
of 2D graphene sheet. That will not lead to misunderstanding
because the latter is a particular case of the former when the
number of layers equals one.} as a function of its orientation.
The hexagonal lattice of the graphene has a three--fold symmetry
and allows me to consider only a sector of the full plane
between 0 and $2\pi/6$.

The function of specific facet energy, $F(\omega)=E(\omega)/L$,
is a periodic function of the orientation angle, $\omega$. At
the fixed $\omega$ it has a weak dependence on the facet length,
$L$, (in 3D case it could depend on the length of the sheet edge
or on the free surface area, which is defined by the specific
type of misorientation). This additional dependence is fully
neglected if considering the continuous edge, owing to full
neglect of the atomic structure. A microscopic view is that at
the scale of interatomic distance, $b$, one can not choose an
arbitrary orientation of the edge. The natural approach to the
description of free misoriented facet is to consider a
number of one--atom kinks distributed along the edge so that
their density depends on the (continuous) orientation angle.

The orientational dependence of the facet energy appears to be
important for the edge zipping model of the single--wall
nanotube nucleation\cite{mrs01}. In order to understand the
experimental fact that synthesized SWNTs had indices
[n,n$\pm\delta$n] close to armchair helicity \footnote{To mark a
SWNT I follow the conventional index agreement proposed in
Ref.\cite{dress}.} the zipping of the graphite edges with
{\em different orientations} has to be considered.

\section{Chirality dependent formation energy}
\label{sec:chir}

The surface
energy of free graphene facet depends on its orientation via
the energy of dangling bonds (DBs).
In this context, the term {\em free facet} means that there is no
special reconstruction of the edge bonds. Of course, the
reconstruction can change the facet energy substantially. I
consider some type of the 3D facet reconstruction in
Ref.\cite{unp}. The reconstruction of the 2D facet is somewhat
more complicated as the system possesses less degree of freedom.
It will be discussed at length elsewhere.

\begin{figure}[h]
\centerline{ \psfig{figure=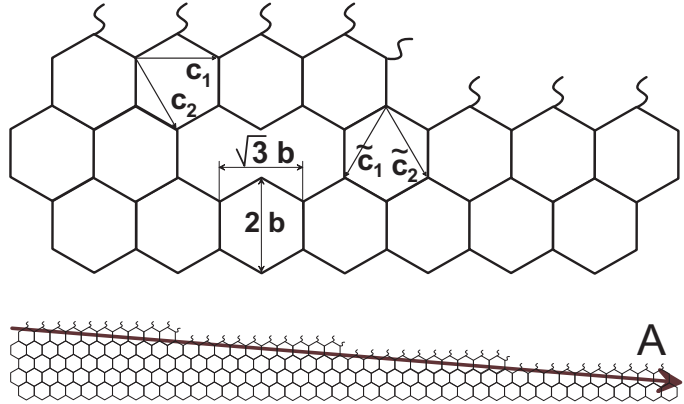,width=9cm}}
\vskip .2cm
\caption{\label{fig:edge}
Misoriented edge of the graphene layer.
The exact direction of the 2D-crystal-face, $\vec A$, is defined
as an envelope along the kinked edge. Any
edge, being different from zigzag and armchair basal types, is,
at the atomic scale, a series of kinks and steps of these basal
faces. Upper: Details of the single kink at the zigzag edge and
the unit cell geometry of the graphene.} \end{figure}

The microscopic picture of
the free edge is presented in Fig.\ref{fig:edge} (upper). Along
the graphene edge the dangling bonds are schematically drawn.
The orientation of the main edge is chosen along [$2{\bar 1}0$]
zigzag direction \footnote{There are multiple possible choice of
the unit vector diads; the second (tilde) set, shown in the
Figure 1 on the right, gives the edge symbol [${\bar 1}{\bar
1}0$]. Both choices are equivalent to [110]}, which is defined
respectively to the basis lattice unit vectors $\vec c_1$ and
$\vec c_2$ (the third unit vector $\vec c_3$ is normal to the
graphite plane, so the corresponding lattice number is always
equal to 0 for graphene). The density of DBs increases at such
discontinuities of the edge as the kink shown in the
Figure\ref{fig:edge}.

%\newpage

The total energy of DBs constitutes the free surface energy:
\begin{equation}
E_{\rm edge}=N_{\rm layers} \,
\nu(\omega) \, L  \, E_b ,
\label{surf1}
\end{equation}
here the edge density of DBs, $\nu(\omega)$, the edge length,
$L$, and the number of layers, $N_{\rm layers}$, give the total
number of DBs. $E_b\simeq 2.355$ eV is the single DB
energy\cite{pla}. I do not consider here any
passivation/contamination of the graphite surface, which could
change $E_b$.

\begin{figure}[h]
\centerline{ \psfig{figure=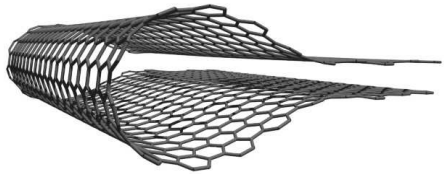,width=6cm}}
\vskip .2cm
\caption{\label{fig:sleeve}
Simulation of the graphene edge zipping (after
[14]).}
%%%\cite{kirchb}).}
\end{figure}

The less the edge density, the more stable the surface. It is
the zigzag face normal to [110] (or symmetry equivalent)
direction\cite{graphite}. The edge, being scrolled and
zipped (as shown in Fig.\ref{fig:sleeve}), will form a
cylindrical sleeve (SWNT nucleus). Its axis is along direction
[${\bar 1}10$] in the graphene plane. This corresponds exactly
to armchair nanotube conformation. I speculate that it gives a
plausible explanation of the preferable formation of {\em armchair
SWNTs} at certain conditions. The direct observation of
graphene scrolling was published in
Ref.\cite{sattler1,sattler2,steps}. It gives also an
experimental support that the stable graphene edge is the zigzag
one.

%\newpage

The edges of zigzag and armchair types
have the minimum and maximum densities of the DBs. It is possible
to expand the DB density around extremum points and
write an interpolation formula in between. Surprising
cusp\footnote{A similar nonmonothonic behavior is known for the
function of the surface energy of perfect crystals for
a century at least
(G.S.Wulff, Z.Krist.Mineral. v. 34, 449, 1901).} is
found for the minimum, where the density changes linearly in
$\omega$, the chirality angle. The latter is a standard measure
of how far does the SWNT axis deviate from the high symmetry
direction, for example, the zigzag direction. Because my edge
structures have 1D geometry close to tubular one (Fig.
\ref{fig:sleeve}), it is natural to characterize their
misorientation with the same chirality angle.

Let me define the edge density through the nanotube--like
indices n,m which are the numbers of unit vectors
constituting the vector $\vec A$ joining the left and right
ends of the considered edge structure [n m 0]
(Fig.\ref{fig:edge}). The DB
density is the number of boundary sites divided by the length of
the boundary:
\begin{equation}
|A|=\sqrt{3}b\sqrt{n^2+n m +m^2},
\label{boundary}
\end{equation}
where $b=|c|/\sqrt{3}\simeq 1.41$ \AA~ is the carbon bond
length. The
infinitesimal small shift of the vector $A=$[n,n] (or [n,0])
defining the armchair (or zigzag) basal edge results in the
change of the edge length and the number of DBs. I expand both
in series on the small parameter $1/n \propto \omega$.
For an armchair edge it reads as:
\begin{equation}
\nu_A=\frac{2n+1}{3\, b\, n\sqrt{1+1/n+1/3n^2}}
\sim \frac{2}{3\, b}\left(
1-\frac{1}{2n}+\frac{1}{2n}-\frac{1}{6n^2}+\frac{3}{8n^2}
-\frac{1}{4n^2}+\dots\right)= \frac{2}{3\, b}\left( 1 -
\frac{1}{36} \omega^2+\dots\right).
\label{dba}
\end{equation}
Naturally,
the first order terms cancel while the second order terms give
the square dependence in the chirality angle, $\omega\propto
\frac{1}{n}$, around the maximum. The coefficient is
$k_A=-\frac{1}{36}$. The behavior of a nearly zigzag edge
density is different. The non--zero expansion terms appear in
the first order:
\begin{equation}
\nu_Z=\frac{n+1}{\sqrt{3}\, b\, n\sqrt{1+1/n+1/n^2}}\sim
\frac{1}{\sqrt{3}\, b}\left(
1+\frac{1}{n}-\frac{1}{2\, n}-\dots\right)
=\frac{1}{\sqrt{3}\, b}\left(
1+\frac{1}{2\sqrt{3}}\omega+\dots\right).
\label{dbz}
\end{equation}
Hence, the linear
dependence coefficient is $s_Z=+\frac{1}{2\sqrt{3}}$ and the
density of DBs has a cusp near the point of the minimum.

An edge, which has the intermediate crystallography index [n m 0]
is nothing but a series of kinks (single atom steps) along
the basal edge of a definite type (Fig.\ref{fig:edge}).

\begin{figure}[h]
\centerline{ \psfig{figure=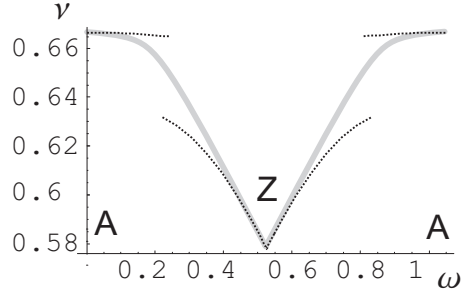,width=6cm}}
\vskip .2cm
\caption{\label{fig:surf}
Dangling bond density of the graphene edge vs. chirality angle.
The surface energy of the graphene fragment is periodic function
of $\omega$ because of $\nu(\omega)$. Dotted lines represent
the approximations at the extremum points given by
Eqs.(3,4).}
\end{figure}

Let me consider the (approximately) circular fragment of
graphene, which is naively expected to deliver the minimum to
the surface energy in a continuum approximation. The actual
surface energy will oscillate along the perimeter with the
period of $\pi/3$ between maxima corresponding to equivalent
armchair edges (Figure \ref{fig:surf}).

%\newpage

At zero temperature, the purely zigzag facing of a 2D--crystal is
expected. At the thermodynamic equilibrium, the zigzag--like
edges are formed and the energetic cost for the vicinal edge is
proportional to $\exp(-\nu L E_b/T)$, where $\nu$ depends on the
orientation (facing angle) linearly, see Eq.(\ref{dbz}).

Let me quantify this.
The energy gain for the
nucleation of the vicinal zigzag edge [n, $\delta$n, 0] is as
follows:
\begin{equation}
\delta E(\omega)=
N_{\rm layers}\,
\delta\nu(\omega)
\, E_b \, L
\simeq
N_{\rm layers} \, L \,
\nu_Z
\, E_b \, s_Z \omega
\simeq
\frac{N_{\rm layers}\, L\omega}{6\, b} \, E_b .
\label{vicin}
\end{equation}
Here I subtract the DB energy of the optimum zigzag edge,
$N_{\rm layers} \, L \, \nu_Z \, E_b$. The chirality equals
$\omega=\delta n/n$.

For the zipping model
of SWNT nucleation one has $N_{\rm layers}=2$,
it yields the following
exponentially small probability of the nucleation
of chiral SWNT with indices
[n, n$\pm\delta$n] at the temperature $T$:
\begin{equation}
w\sim w_o \exp [-\frac{1}{T}\frac{\omega L}{3 b}E_b]=
w_o \displaystyle
e^{\displaystyle -\frac{\displaystyle \omega}{\omega_c}},
\label{prob}
\end{equation}
where $w_o$ is the probability of nucleation of armchair [n,n]
SWNT, $\displaystyle \omega_c=\frac{3b}{L}\frac{T}{E_b}$
is a maximum
chirality at given $L$ and $T$. The higher the temperature, the
shorter the nucleus length. Thus the $\omega_c$ has a maximum
as a function of temperature. Let me estimate $L\simeq 50$\AA~
at $T\simeq 2000$~K. Then the $\omega_c\simeq0.006$
 and the probability to nucleate [10,11] tube is  $10^{-7}$
 smaller than for [10,10] SWNT
%($D\simeq 100$\AA~ at $T\simeq 1000$~K gives
%$\omega_c\simeq0.0015$ and the probability is $10^{-28}$).
%

Let me consider the edge closest to zigzag one. The smallest edge
defect is a couple of kinks (per total nucleus length) with
the energy equals $E_b$. This minimum energy  gain corresponds
to the relative nucleation probability
$w/w_o\sim e^{-E_b/T}\simeq 10^{-6}$ at $T=2000$K.

The calculated probability of chiral SWNT nucleation is too
small to describe chirality scattering of
the experiment. The real growing edge shape is
likely far from equilibrium (ideal straight line). Therefore,
the kinetics of graphene edge growth defines both the length and
the chirality of the nucleus. I did not consider here the
dangling bond passivation. Whence the kink energy is $E_b/2$,
the passivation increases the maximum allowed chirality that
could also explain the experimental scattering of SWNT angles.

%*************************************************
%\eject

\section{Summary}

The microscopic model describing the energy of the edge
formation for different types of graphitic structures is
developed. The most stable edge has a zigzag orientation owing
to the smallest density of the dangling bonds along the facet.
The DB density is calculated depending on the misorientation
angle for the facet of the arbitrary direction.

In respect to proposed earlier
SWNT nucleation mechanism via formation
of a cylindrical sleeve along the
graphene edge, this paper allows to speculate that
the prevalent tubes have almost the armchair
type which is complementary to the most stable zigzag graphene
edge. The prevalent radius of such a sleeve nucleus is
close to the optimum radius $R_m$ of nano--arch
structure considered in \cite{mrs01}. My estimation gives a
lower limit of the uniformity of the SWNT synthesis yield. In
ideal conditions, the novel nucleation mechanism predicts
formation of [11,11] SWNT (and no chirality
scattering is expected) which is best approximation of continual
energetics model to the most abundant [10,10] SWNT observed
experimentally.

Note, that in this paper I did not consider
any kinetics of the formation, that might change the result.


\begin{references}
\bibitem{nanocarbons}
Not pretending to cover all articles, a number of related papers
is listed in \cite{newreview}.
\end{references}
\end{document}